\documentclass[11 pt]{amsart}

\usepackage{amssymb, amsmath}

\usepackage{amsthm}

\newtheorem{theorem}{Theorem}[section]
\newtheorem{problem}[theorem]{Problem}

\newtheorem{lemma}[theorem]{Lemma}
\newtheorem{remark}[theorem]{Remark}
\newtheorem{corollary}[theorem]{Corollary}
\newtheorem{definition}[theorem]{Definition}

\newcommand{\N}{\mathcal{N}}

\begin{document}

\title{Neighborhoods of binary self-dual codes}


\author[C. Hannusch]{Carolin Hannusch}
\author[S.R. Major]{Sándor Roland Major}

\address[C. Hannusch]{Department of Computer Science, Faculty of Informatics, University of Debrecen, email: hannusch.carolin@inf.unideb.hu}
        
\address[S.R. Major]{Department of Information Technology, Faculty of Informatics, University of Debrecen, email: major.sandor@inf.unideb.hu}

\maketitle

\begin{abstract}
In this paper, we introduce and investigate the neighborhood of binary self-dual codes. We prove that there is no better Type I code than the best Type II code of the same length. Further, we give some new necessary conditions for the existence of a singly-even $(56,28,12)$-code and a doubly-even $(72,36,16)$-code. 
\end{abstract}

\section{Introduction}\label{intro}

Neighbors of self-dual codes were first investigated in \cite{conway1990new} and \cite{brualdi1991weight}. Later, neighbors were used to find extremal $(64,32,12)$ codes in \cite{chigira2007extremal} and to find new codes of length $68$ in \cite{dougherty2019bordered}. Recently, the graph of neighboring codes was investigated \cite{dougherty2022neighbor}. In the current paper, we introduce the definition of a neighborhood of binary self-dual codes. The paper is organized as follows: In Section \ref{preliminaries} we mention the main definitions and preliminary results which were essential for our work. Section \ref{auxiliary} contains some auxiliary results. In Section \ref{neighbors} we investigate the relation of neighboring self-dual codes and introduce the neighborhood of self-dual codes. In Section \ref{unique} we show that equivalent codes can have different neighborhoods. Finally, in Section \ref{conclusion} we give some research problems.

\section{Preliminaries}\label{preliminaries}

Let $\mathbb{F}_2$ denote the finite field of two elements and let $n$ be a positive integer. Then a subspace of $\mathbb{F}_2^n$ is called a \textit{binary linear code}. We denote a linear code by $C.$ Then its \textit{dual code} $C^{\bot}$ is defined as $$C^{\bot} = \{x \in \mathbb{F}_2^n | x\cdot c = 0 \mbox{  } \forall c \in C\},$$ where $\cdot$ denotes the usual scalar product of two vectors. A code is called \textit{self-orthogonal} if $C\subseteq C^{\bot}$ and \textit{self-dual} if $C = C^{\bot}.$ It is a well known fact that the dimension of a self-dual code of length $n$ is $\frac{n}{2}.$ 

The \textit{weight} of a codeword $c\in C$ is the number of its nonzero coordinates. The \textit{minimum weight} (or \textit{minimum distance}) of a code $C$ is the smallest nonzero weight of its codewords. We denote the minimum distance of $C$ by $d(C).$ If a code $C$ is a $k$-dimensional subspace of $\mathbb{F}_2^n$ with minimum distance $d,$ then we say that $C$ is an $(n,k,d)$-code. 

Self-dual binary codes can be classified into Type I and Type II codes \cite{joyner2011selected}. A self-orthogonal binary code $C$ is said to be \textit{doubly-even} if all of its codewords have weight divisible by $4.$ If $C$ has a codeword of weight not divisible by $4,$ then $C$ is \textit{singly-even} code \cite{huffman2010fundamentals}. Singly-even self-dual codes are called \textit{Type I} codes and doubly-even self-dual codes are called \textit{Type II} codes \cite{joyner2011selected}.  It is well known that Type II codes only exist for length divisible by $8$ \cite{huffman2005classification}. 

We know that the minimum distance of a binary self-dual code is bounded by $d(C) \leq 2\lfloor \frac{n}{8}\rfloor +2$ if $C$ is a Type I code and $d(C) \leq 4\lfloor \frac{n}{24}\rfloor +4$ if $C$ is a Type II code (\cite{conway1990new}, \cite{huffman2005classification}). If a code reaches equality in this bound, then the code is called extremal. Especially for larger codelengths it seems that equality cannot be reached in many cases. Codes with the highest possible minimum distance are called optimal codes. The search for extremal and optimal binary codes is a difficult task for many codelengths $n$ and several researchers have contributed to this theory; see e.g., \cite{kim2001new}, \cite{bouyuklieva2002some}. There are still many open problems about extremal and optimal binary codes \cite{joyner2011selected}.

\section{Auxiliary results}\label{auxiliary}

Let $a$ and $b$ be two codewords (i.e.~binary vectors) of the same length. We denote the numbers of coordinates, which are $1$ in both codewords by $\mu(a,b),$ i.e.~
$$\mu(a,b)=\#\{i\mid a[i]=b[i]=1, i\in \{1,\ldots,n\}\}.$$ The weight of the sum of two codewords is the following \cite{brualdi1991weight} $$w(a+b) = w(a)+w(b) - 2\mu(a,b).$$

\begin{lemma}\label{mu-addition}
	Let $a,b,c$ be codewords of the same length. Then 
	$$\mu(a+b,c) = \mu(b,c) + \mu(a,b+c) - \mu(a,b).$$
\end{lemma}

\begin{proof}
	Using the following equations in the given order, the proposition can be directly shown:
	\begin{enumerate}
		\item $w(a+b+c)=w(a+b)+w(c) -2 \mu(a+b,c)$
		\item $w(a+b)=w(a)+w(b)-2\mu(a,b)$
		\item $w(a+b+c)=w(a)+w(b+c) -2 \mu(a,b+c)$
		\item $w(b+c)=w(b)+w(c)-2\mu(b,c)$ 
	\end{enumerate} 
\end{proof}

\section{Neighbors and neighborhoods}\label{neighbors}

In \cite{brualdi1991weight} the definition of neighbors among self-dual codes was introduced as follows. 

\begin{definition}
	Two self-dual codes of length $n$ are called neighbors, provided their intersection is a code of dimension $\frac{n}{2}-1.$
\end{definition}

  It is well known that a Type I code has a maximal doubly-even subcode of codimension $1.$ Throughout the paper, we will denote this maximal doubly-even subcode by $C_{max},$ further we denote the all-1 codeword by $\textbf{1}.$ If $C$ is a code of Type I of length divisible by $8,$ then $\textbf{1}\in C.$ 

We know by \cite{conway1990new} and \cite{brualdi1991weight} that if $C$ is Type I, then it has two Type II neighbors. This implies that all singly-even codewords of $C_{max}^{\perp}$ have weight $w,$ where $d\leq w \leq n-d.$ 

Investigating the neighbors of self-dual codes leads us to some interesting facts about the relation of neighboring codes. We find out that codes with the same maximal subcode have a special relation to each other. Therefore we come up to the definition of a neighborhood for binary self-dual codes. 

We extend the definition of neighbors to a set of codes, which are pairwise neighbors. We will call this set of codes a neighborhood.

\begin{definition}
	Let $C_{max}$ be a self-orthogonal $(n,k-1,d)$-code, where $k=2n.$ Then there exist three self-dual codes $C_1, C_2$ and $C_3,$ which contain $C_{max}$ as a maximal subcode of codimension $1,$ i.e.~they are pairwise neighbors. Then we say, that $\{C_1, C_2,C_3\}$ is a neighborhood of codes. We will denote this set by $\N.$ 
\end{definition}

\begin{remark}
	The neighborhood $\N$ of a self-dual code always consists of three codes, one of Type I and two of Type II. 
\end{remark}

In the following, we investigate the minimum distances of three members of a neighborhood $\N$ and their relations. First, it turns out that if the Type I member of a neighborhood $\N$ has minimum distance $2,$ then the minimum distance of the two Type II members coincide.

\begin{theorem}
	Let $\N$ be a neighborhood of binary self-dual codes of length divisible by $8.$ If the singly-even member of $\N$ has minimum distance $2,$ then the minimum distance of the doubly-even members coincide. 
\end{theorem}

\begin{proof}  We denote the members of $\N$ by $C_1, C_2$ and $C_3.$ Their coinciding doubly-even subcode will be denoted by $C_{max}.$ Then $C_1=\langle C_{max}, \gamma_1 \rangle,$ $C_2=\langle C_{max}, \gamma_2 \rangle$ and $C_3=\langle C_{max}, \gamma_1 + \gamma_2 \rangle$ for suitable $\gamma_1,\gamma_2\in C_{max}^{\bot}.$ Furthermore, we assume that $w(\gamma_2)\equiv w(\gamma_1+\gamma_2) \equiv 0 \mod 4.$ Since $\mu(\gamma_2,\gamma_1+\gamma_2)\equiv 1 \mod 2,$ we have $w(\gamma_1)\equiv 2 \mod 4.$ Thus $C_1$ is a singly-even code and $C_2, C_3$ are doubly-even codes. By assumption, we have $d(C_1)=2$ and thus we can choose $\gamma_1$ such that $w(\gamma_1)=2.$ Let us denote the minimum distance of $C_2$ by $d.$ Then we have for all $c\in C_{max}$ $$d\leq w(c+\gamma_2) \leq n-d.$$ We know $$w(c+\gamma_1+\gamma_2) = w(c+\gamma_2) + w(\gamma_1) - 2\mu(c+\gamma_2,\gamma_1).$$ Since $c+\gamma_2$ and $\gamma_1$ cannot be orthogonal, we have $\mu(c+\gamma_2,\gamma_1)=1.$ Thus $w(c+\gamma_1+\gamma_2)=w(c+\gamma_2)$ for all $c\in C_{max},$ which implies $d(C_3)=d.$
\end{proof}

\begin{theorem}\label{nobettertypei}
	Let $\N=\{C_1,C_2,C_3\}$ be a neighborhood of self-dual codes of length divisible by $8.$ We assume that $C_1$ is Type I and $C_2$ and $C_3$ are Type II. Then $d(C_1)\leq \max\{d(C_2),d(C_3)\}.$
\end{theorem}

\begin{proof}
	For technical reasons and to keep the proof as simplest as possible, we assume now that $C_1=\langle C_{max}, \gamma_1 + \gamma_2\rangle$ is a singly-even code and $C_2=\langle C_{max}, \gamma_1\rangle$, $C_3=\langle C_{max}, \gamma_2\rangle$ are doubly-even codes.
	
	We assume indirectly that $d=d(C_1)> \max\{d(C_2),d(C_3)\}.$ 
	Then for any $c_i\in C_{max}$ we have $$d\leq w(c_i)\leq n-d$$ and $$d\leq w(c_i+\gamma_1 + \gamma_2)\leq n-d \Leftrightarrow$$
	$$d\leq w(c_i+\gamma_1) + w(\gamma_2) - 2\mu(c_i+\gamma_1,\gamma_2) \leq n-d \Leftrightarrow$$
	$$\max\{d(C_2),d(C_3)\} < d \leq w(c_i) + w(\gamma_1) + w(\gamma_2) -2\mu(c_i,\gamma_1) - 2\mu(c_i+\gamma_1,\gamma_2)  \leq n-d $$
	$$ \Leftrightarrow 2\mu(c_i,\gamma_1) + 2\mu(c_i+\gamma_1,\gamma_2) \leq w(\gamma_1) + w(\gamma_2)$$
	Then using Lemma \ref{mu-addition} in the lefthand side we get
	$$2\mu(c_i,\gamma_1+\gamma_2) + 2\mu(\gamma_1,\gamma_2) \leq w(\gamma_1)+w(\gamma_2).$$
	Since $\mu(c_i,\gamma_1+\gamma_2)\equiv 0 \mod 2$ and $\mu(\gamma_1,\gamma_2)\equiv 1 \mod 2$ we have 
	$$2\mu(c_i,\gamma_1+\gamma_2) + 2\mu(\gamma_1,\gamma_2) < w(\gamma_1)+w(\gamma_2).$$
	Thus $$2\mu(c_i,\gamma_1+\gamma_2)  < w(\gamma_1)+w(\gamma_2) - 2\mu(\gamma_1,\gamma_2) = w(\gamma_1+\gamma_2).$$
	Thus $w(c_i+\gamma_1+\gamma_2) > w(c_i) + w(\gamma_1+\gamma_2) - w(\gamma_1+\gamma_2),$ therefore  $w(c_i+\gamma_1+\gamma_2) > w(c_i)$ for each $c_i\in C_{max}$ which is a contradiction, since if $w(c_i)=n-d,$ then $w(c_i+\gamma_1+\gamma_2) > n-d,$ which implies $d(C_1) < d,$ since $\textbf{1}\in C_1.$
\end{proof}

\begin{corollary}\label{nobetter}
	There is no better Type I code than the best possible Type II code of the same length. 
\end{corollary}

This answers a question of Rains and Sloane \cite{rains1998self} and Conway and Sloane (\cite{conway1990new}, p. 1321, Open question 1) who asked what is the smallest $n,$ such that a Type I code of length $n$ is better than the best Type II code of the same length. Using the neighborhood approach we can see that such a code cannot exist. This shows that the neighborhood approach enables us to see binary self-dual codes in a different way and thus to see relations and properties which were not obvious before.  

\section{The existence of certain self-dual codes}

The existence of a Type I $(56,28,12)$-code is an open question \cite{dougherty2015open}. It is known that Type II $(56,28,12)$-codes exist \cite{harada2018singly}, \cite{yorgov1987method}, \cite{bussemaker1989new}. Considering our previous results, we come to the following fact.

\begin{corollary}
	If there exists a Type I $(56,28,12)$-code, then it is a neighbor of a Type II $(56,28,12)$-code, since $12$ is the best possible minimum distance for a self-dual code of length $56.$
\end{corollary}

Another unsolved question is the existence of a doubly-even $(72,36,16)$-code \cite{dougherty2015open}. By Corollary \ref{nobetter} we can reduce the search of a Type II $(72,36,16)$-code to the search of a Type I $(72,36,14)$-code.

\begin{corollary}
	If there exists a Type I $(72,36,14)$-code, then it has a doubly-even neighbor with greater minimum distance, i.e.~it has a Type II $(72,36,16)$-code as neighbor.
\end{corollary}

\section{Neighborhood is not unique}\label{unique}

The neighborhood of a self-dual code is not unique for permutation equivalent codes. Equivalent codes can be contained in neighborhoods whose members have different properties (minimum distance). For example, the well-known $(24,12,8)$ Golay code is unique up to permutation equivalence, but it is contained in (at least) two distinct neighborhoods. The first neighborhood is $\N_1 = \{C_1, C_2, C_3\},$ where $C_1$ and $C_2$ are two permutation equivalent $(24,12,8)$-codes and $C_3$ is a $(24,12,2)$-code. The second neighborhood is $\N_2 = \{C_4, C_5, C_6\},$ where $C_4$ is a $(24,12,6)$-code, $C_5$ is a $(24,12,4)$-code and $C_6$ is a $(24,12,8)$-code. Generator matrices for $C_1, C_2, C_3, C_4, C_5$ and $C_6$ are the following matrices $G_1, G_2, G_3, G_4, G_5$ and $G_6,$ respectively.

\tiny

\[
G_1 = \left( \begin{array}{c}
	 1 0 0 0 0 0 0 0 0 0 0 1 1 1 1 1 1 1 1 1 1 0 0 1 \\
	 0 1 0 0 0 0 0 0 0 0 0 0 1 1 1 1 1 1 0 0 0 1 0 0 \\
	 0 0 1 0 0 0 0 0 0 0 0 0 1 1 1 0 0 0 1 1 1 1 0 0 \\
	 0 0 0 1 0 0 0 0 0 0 0 0 1 0 1 1 1 0 1 0 1 0 1 0 \\
	 0 0 0 0 1 0 0 0 0 0 0 0 0 1 1 1 0 1 0 1 1 0 1 0 \\
	 0 0 0 0 0 1 0 0 0 0 0 1 1 0 1 0 0 0 0 1 1 0 1 1 \\
	 0 0 0 0 0 0 1 0 0 0 0 1 0 1 1 0 0 1 0 0 1 1 0 1 \\
	 0 0 0 0 0 0 0 1 0 0 0 1 1 1 0 0 1 0 0 1 0 1 0 1 \\
	 0 0 0 0 0 0 0 0 1 0 0 1 0 0 0 1 0 1 1 0 1 0 1 1 \\
	 0 0 0 0 0 0 0 0 0 1 0 1 1 0 1 1 0 0 1 0 0 1 0 1 \\
	 0 0 0 0 0 0 0 0 0 0 1 1 0 1 1 1 1 0 0 0 0 0 1 1 \\
	 0 0 0 0 0 0 0 0 0 0 0 0 0 0 1 0 1 1 1 1 0 1 1 1 
	 \end{array} \right)
G_2 = \left( \begin{array}{c}
	1 0 0 0 0 0 0 0 0 0 0 1 1 1 1 1 1 1 1 1 1 0 0 1 \\
	0 1 0 0 0 0 0 0 0 0 0 0 1 1 1 1 1 1 0 0 0 1 0 0 \\
	0 0 1 0 0 0 0 0 0 0 0 0 1 1 1 0 0 0 1 1 1 1 0 0 \\
	0 0 0 1 0 0 0 0 0 0 0 0 1 0 1 1 1 0 1 0 1 0 1 0 \\
	0 0 0 0 1 0 0 0 0 0 0 0 0 1 1 1 0 1 0 1 1 0 1 0 \\
	0 0 0 0 0 1 0 0 0 0 0 1 1 0 1 0 0 0 0 1 1 0 1 1 \\
	0 0 0 0 0 0 1 0 0 0 0 1 0 1 1 0 0 1 0 0 1 1 0 1 \\
	0 0 0 0 0 0 0 1 0 0 0 1 1 1 0 0 1 0 0 1 0 1 0 1 \\
	0 0 0 0 0 0 0 0 1 0 0 1 0 0 0 1 0 1 1 0 1 0 1 1 \\
	0 0 0 0 0 0 0 0 0 1 0 1 1 0 1 1 0 0 1 0 0 1 0 1 \\
	0 0 0 0 0 0 0 0 0 0 1 1 0 1 1 1 1 0 0 0 0 0 1 1 \\
	0 0 0 0 0 0 0 0 0 0 0 1 0 0 1 0 1 1 1 1 0 1 1 0 
 \end{array}\right)
\]

\[
G_3 = \left( \begin{array}{c}
1 0 0 0 0 0 0 0 0 0 0 1 1 1 1 1 1 1 1 1 1 0 0 1 \\
0 1 0 0 0 0 0 0 0 0 0 0 1 1 1 1 1 1 0 0 0 1 0 0 \\
0 0 1 0 0 0 0 0 0 0 0 0 1 1 1 0 0 0 1 1 1 1 0 0 \\
0 0 0 1 0 0 0 0 0 0 0 0 1 0 1 1 1 0 1 0 1 0 1 0 \\
0 0 0 0 1 0 0 0 0 0 0 0 0 1 1 1 0 1 0 1 1 0 1 0 \\
0 0 0 0 0 1 0 0 0 0 0 1 1 0 1 0 0 0 0 1 1 0 1 1 \\
0 0 0 0 0 0 1 0 0 0 0 1 0 1 1 0 0 1 0 0 1 1 0 1 \\
0 0 0 0 0 0 0 1 0 0 0 1 1 1 0 0 1 0 0 1 0 1 0 1 \\
0 0 0 0 0 0 0 0 1 0 0 1 0 0 0 1 0 1 1 0 1 0 1 1 \\
0 0 0 0 0 0 0 0 0 1 0 1 1 0 1 1 0 0 1 0 0 1 0 1 \\
0 0 0 0 0 0 0 0 0 0 1 1 0 1 1 1 1 0 0 0 0 0 1 1 \\
0 0 0 0 0 0 0 0 0 0 0 1 0 0 0 0 0 0 0 0 0 0 0 1 
\end{array} \right)
G_4 = \left( \begin{array}{c}
1 0 0 0 0 0 0 0 0 0 0 0 1 1 1 1 1 1 1 1 1 1 1 0 \\
0 1 0 0 0 0 0 0 0 0 0 1 0 1 0 1 1 0 1 1 1 0 0 0 \\
0 0 1 0 0 0 0 0 0 0 0 1 0 1 0 1 1 1 0 0 0 1 1 0 \\
0 0 0 1 0 0 0 0 0 0 0 1 0 0 1 0 1 0 1 1 0 1 1 0 \\
0 0 0 0 1 0 0 0 0 0 0 0 0 1 1 1 1 0 0 1 0 0 1 1 \\
0 0 0 0 0 1 0 0 0 0 0 0 0 1 0 0 1 1 1 1 0 1 0 1 \\
0 0 0 0 0 0 1 0 0 0 0 0 0 0 1 0 1 1 0 0 1 1 1 1 \\
0 0 0 0 0 0 0 1 0 0 0 1 1 1 1 0 1 0 0 0 1 0 1 0 \\
0 0 0 0 0 0 0 0 1 0 0 1 0 1 1 0 0 1 1 0 1 1 0 0 \\
0 0 0 0 0 0 0 0 0 1 0 1 1 0 0 0 1 1 0 1 1 1 0 0 \\
0 0 0 0 0 0 0 0 0 0 1 1 0 0 1 1 0 1 0 1 1 0 1 0 \\
0 0 0 0 0 0 0 0 0 0 0 1 1 0 1 0 0 1 0 0 1 0 0 1 	
\end{array}\right)
\]

\[
G_5 = \left( \begin{array}{c}
1 0 0 0 0 0 0 0 0 0 0 0 1 1 1 1 1 1 1 1 1 1 1 0 \\
0 1 0 0 0 0 0 0 0 0 0 1 0 1 0 1 1 0 1 1 1 0 0 0 \\
0 0 1 0 0 0 0 0 0 0 0 1 0 1 0 1 1 1 0 0 0 1 1 0 \\
0 0 0 1 0 0 0 0 0 0 0 1 0 0 1 0 1 0 1 1 0 1 1 0 \\
0 0 0 0 1 0 0 0 0 0 0 0 0 1 1 1 1 0 0 1 0 0 1 1 \\
0 0 0 0 0 1 0 0 0 0 0 0 0 1 0 0 1 1 1 1 0 1 0 1 \\
0 0 0 0 0 0 1 0 0 0 0 0 0 0 1 0 1 1 0 0 1 1 1 1 \\
0 0 0 0 0 0 0 1 0 0 0 1 1 1 1 0 1 0 0 0 1 0 1 0 \\
0 0 0 0 0 0 0 0 1 0 0 1 0 1 1 0 0 1 1 0 1 1 0 0 \\
0 0 0 0 0 0 0 0 0 1 0 1 1 0 0 0 1 1 0 1 1 1 0 0 \\
0 0 0 0 0 0 0 0 0 0 1 1 0 0 1 1 0 1 0 1 1 0 1 0 \\
0 0 0 0 0 0 0 0 0 0 0 0 0 0 0 1 1 0 1 0 1 0 0 0 
\end{array} \right)
G_6 = \left( \begin{array}{c}
1 0 0 0 0 0 0 0 0 0 0 0 1 1 1 1 1 1 1 1 1 1 1 0 \\
0 1 0 0 0 0 0 0 0 0 0 1 0 1 0 1 1 0 1 1 1 0 0 0 \\
0 0 1 0 0 0 0 0 0 0 0 1 0 1 0 1 1 1 0 0 0 1 1 0 \\
0 0 0 1 0 0 0 0 0 0 0 1 0 0 1 0 1 0 1 1 0 1 1 0 \\
0 0 0 0 1 0 0 0 0 0 0 0 0 1 1 1 1 0 0 1 0 0 1 1 \\
0 0 0 0 0 1 0 0 0 0 0 0 0 1 0 0 1 1 1 1 0 1 0 1 \\
0 0 0 0 0 0 1 0 0 0 0 0 0 0 1 0 1 1 0 0 1 1 1 1 \\
0 0 0 0 0 0 0 1 0 0 0 1 1 1 1 0 1 0 0 0 1 0 1 0 \\
0 0 0 0 0 0 0 0 1 0 0 1 0 1 1 0 0 1 1 0 1 1 0 0 \\
0 0 0 0 0 0 0 0 0 1 0 1 1 0 0 0 1 1 0 1 1 1 0 0 \\
0 0 0 0 0 0 0 0 0 0 1 1 0 0 1 1 0 1 0 1 1 0 1 0 \\
0 0 0 0 0 0 0 0 0 0 0 1 1 0 1 1 1 1 1 0 0 0 0 1 
\end{array} \right)
\]

\normalsize

\bigskip

\begin{remark}
	If $C$ is a Type I code, then its maximal doubly-even subcode is unique, thus its neighborhood is unique as well.
\end{remark}

\section{Conclusion and further research}\label{conclusion}

We are convinced that the investigation of binary self-dual codes through their neighborhoods opens new possibilities as well as in finding codes whose existence is not known yet, as well as to understand the relations between self-dual codes better. 

By Theorem \ref{nobettertypei} we know that the minimum distance of a singly-even self-dual code cannot be greater than the minimum distance of its best doubly-even neighbor, but it can be equal.

	There are three non-equivalent $(32,16,8)$ Type I codes \cite{bouyuklieva2012singly}. All of them have a doubly-even neighbor with minimum distance $8,$ which we computed by the program package TORCH \cite{hannusch2022}.

Therefore the following questions arise, whose solutions may help to solve open questions like the existence of a Type I $(56,28,12)$-code or a Type II $(72,36,16)$-code.

\begin{problem} Is there a condition for the codelength $n,$ such that the minimum distance of the best Type II code coincides with the minimum distance of its Type I neighbor?
\end{problem}

\begin{problem} Given a neighborhood $\N = \{C_1,C_2,C_3\}$ is it possible that all three minimum distances coincide, i.e.~$d(C_1) = d(C_2) =d(C_3)$? 
\end{problem}



\bibliographystyle{elsarticle-num} 
\bibliography{bibcodingtheory.bib}





\end{document}